\documentclass[reprint,superscriptaddress,showpacs,nofootinbib,aps,pra]{revtex4-1}
\usepackage{amssymb}
\usepackage{amsmath}
\usepackage{graphicx}
\usepackage{dcolumn}
\usepackage{bm}
\usepackage[utf8]{inputenc}
\usepackage[frenchb]{babel}
\usepackage{braket}

\begin{document}

\title{Effective interactions in a quantum Bose-Bose mixture}

\author{O. I. Utesov}
\affiliation{National Research Center "Kurchatov Institute" B.P.\ Konstantinov Petersburg Nuclear Physics Institute, Gatchina 188300, Russia}

\author{M. I. Baglay}
\affiliation{ITMO University, St. Petersburg 197101, Russia} 

\author{S. V. Andreev}
\email[Electronic adress: ]{Serguey.Andreev@gmail.com}
\affiliation{ITMO University, St. Petersburg 197101, Russia}
\date{\today}

\begin{abstract}
We generalize the Beliaev diagrammatic theory of an interacting spinless Bose-Einstein condensate to the case of a binary mixture. We derive a set of coupled Dyson equations and find analytically the Green's functions of the system. The elementary excitation spectrum consists of two branches, one of which takes the characteristic parabolic form $\omega\propto p^2$ in the limit of a spin-independent interaction. We observe renormalization of the magnon mass and the spin-wave velocity due to the Andreev-Bashkin entrainment effect. For a 3D weakly-interacting gas the spectrum can be obtained by applying the Bogoliubov transformation to a second-quantized Hamiltonian in which the microscopic two-body potentials in each channel are replaced by the corresponding off-shell scattering amplitudes. The superfluid drag density can be calculated by considering a mixture of phonons and magnons interacting via the effective potentials. We show that this problem is identical to the second-order perturbative treatment of a Bose polaron. In 2D the drag contributes to the magnon dispersion already in the first approximation. Our consideration provides a basis for systematic study of emergent phases in quantum degenerate Bose-Bose mixtures.
\end{abstract}

\maketitle

\preprint{APS/123-QED}

\section{Introduction}

The effective interaction is one of the most insightful concepts in the theoretical many-body physics. Correlations between the particles forming the media change the magnitude and may even transform the shape of the microscopic two-body interaction potential. Such modifications become especially profound in the quantum regime. A textbook example is a degenerate electron gas in a lattice of positively charged ions. Renormalization of the Coulomb repulsion due to polarization of the media yields a dipolar-like (pseudo-)potential with a cosine modulation \cite{Fetter}. Experimental manifestation of this effect is known as Friedel oscillations \cite{Friedel}. 

In the case of bosons the physics is further enriched by the presence of a condensate at absolute zero temperature. Quantum scattering of the matter waves in the condensate can be promoted onto the macroscopic scale, which gives birth to new collective states of matter. Predicted in early 70-s the "coherent crystals" \cite{Kirzhnits, Nepomnyashchii} with possible supersolid properties now surface in ultra-cold dipolar gases \cite{Rosenzweig, Rotons, FragmentedSS}. In contrast to the familiar Wigner crystals \cite{Wigner}, crystallization of a Bose gas occurs with increase of the density $n$ and the unit cell can accommodate a macroscopically large amount of particles. In the mean-field picture formation of a supersolid can be described in terms of an effective interaction potential which has negative Fourier components in the vicinity of some finite momentum transfer $\bm k_0$ which satisfies $k_0 \ll n^{1/d}$, where $d$ is the dimension of the space. In the dilute limit existence of such feature for a generic condensate characterized by dipolar repulsion at large interparticle distances has been proven on the basis of the Beliaev diagrammatic approach \cite{RotonSpectrum, 1Dscattering}. 

The Beliaev prescription for a scalar Bose-Einstein condensate consists in replacement of the actual microscopic potential by the off-shell scattering amplitude for two particles in a vacuum \cite{Beliaev}. Negative momentum-dependent correction to the scattering amplitude of dipoles was shown to come from large distances (on the order of the thermal de Broglie distances), where the scattering is governed entirely by the repulsive dipolar tail \cite{DipolarScattering, 1Dscattering}. In order to make this contribution comparable with the contact part two pathways has been explored in ultra-cold atomic systems. 

First, one can use the so-called pancake geometry to allow alignement of the dipoles head-to-tail at short distances \cite{PancakeRotons, Review}. Initially prepared in a uniform state the system collapses into a regular pattern of drops after a quench of the Feshbach-resonant part of the scattering length to its background value \cite{Rosenzweig}. There is, however, no mutual coherence between the drops and their shape is strongly elongated in the transverse direction \cite{Filaments, Pfau}. These two factors make the supersolid scenario unlikely here. The physics of the drops appeared to be interesting on its own right because of the role played by quantum fluctuations in their stabilization \cite{ErbiumDrop, DysprosiumDrop, DipolarDropletsScience} (see also below).

The second idea, put forward in \cite{DiluteSupersolid}, is to use ultra-cold polar molecules in the bilayer geometry with tunneling \cite{PolarMolecules}. The tunneling makes the two-component system effectively behave as a 2D scalar gas with vanishing contact part of the effective interaction and controllable three-body repulsive forces \cite{Petrov2014}. The latter ensures the stability of a crystalline structure, which in this case indeed can be regarded as a true supersolid state. However, experimental realization of this model is challenging since the tunnelling would open a channel for three-body losses of the molecules \cite{Will}.

A promising contribution into the field has come from the semiconductor physics. As has been pointed out recently \cite{FragmentedSS}, the 2002 observation of a regular structure in the photoluminescence pattern of dipolar excitons in quantum wells (QW's) \cite{MOES} may hint toward a form of the coherent crystal. A surprising rarity of the phenomenon has been attributed to the specifics of the exciton-exciton interaction potential \cite{1Dscattering}. Interaction of two excitons with opposite spins can admit a shape resonance, which provides an efficient tool to tune the contact part of the scattering amplitude. The dipoles cannot leave the QW plane and stability of the system in the supersolid phase is guaranteed by formation of bosonic dimers (biexcitons) characterized by strong repulsion. A minimal model which allows one to describe the transition to the supersolid of dimers is a two-species dilute Bose gas with a resonant interspecies interaction \cite{ResonantPairing}.

Besides, studies of two-species Bose mixtures are now gaining momentum due to possibility of revealing beyond mean-field effects in the ultra-dilute regime. Thus, following the original proposal \cite{Petrov2015}, quantum droplets have been realized in atomic samples \cite{BoseMixture}. The very existence of such objects is due to quantum fluctuations. Experimental studies and numerical modelling of these states are guided by analytical perturbative expansion of effective low-energy Hamiltonians \cite{Petrov2015, Petrov2016, PetrovNature}.

These recent theoretical ideas and experimental results indicate a need for an extension of the Beliaev approach to a binary mixture of bosons. A challenging question is interference of different channels in a many-body scattering sequence. It is not obvious \textit{a priori} that the interaction in a mixed condensate can be described in terms of independent two-particle scattering processes.

In this paper we give a generic analytical solution of the problem. We derive a set of coupled Dyson equations and find the Green's functions of the system by using a specific spinor representation. The elementary excitation spectrum consists of two branches, one of which takes the characteristic parabolic form $\omega\propto p^2$ in the limit of spin-independent interactions. To the lowest order in the density parameter 
\begin{equation}
\beta=\sqrt{nR_e^d},
\end{equation}
where $R_e$ is the characteristic range of the microscopic interaction, the diagrams for the self-energy parts decouple into a set of independent ladders. This yields three effective potentials expressed via the corresponding scattering amplitudes. In the case of 3D geometry, these potentials can be used to construct an effective Hamiltonian suitable for the perturbative expansion. The quantum interference of the channels manifests itself in renormalization of the magnon mass and the spin-wave velocity revealing the Andreev-Bashkin entrainment effect \cite{AB}. This feature escapes the standard hydrodynamic approach where the Fourier transform of some phenomenological potential is used to describe the normal modes in terms of small-amplitude oscillations \cite{ResonantPairing, Petrov2015, Petrov2016, Goldstein, Berman, Alexandrov, Eckardt}. For a 3D weakly-interacting gas the drag density can be obtained by considering interaction of magnons with the Bogoliubov phonon modes. We show that this problem is identical to the second-order perturbation theory of a Bose polaron developed in \cite{Christensen}. We exploit this fruitful analogy to speculate on possible transition to a \textit{magnon crystal} in the strongly-interacting regime. For weak interactions in 2D the drag contributes to the dispersion already in the first order in $\beta$. This reflects the enhanced role of quantum fluctuations in low dimensions. On the basis of our findings, we expect the entrainment to cause an increasing departure of the quantum correction to the energy of the mixture from the predictions \cite{Petrov2015, Petrov2016}.   

\section{The model}
\begin{figure*}[t]
\centering{
\includegraphics[width=1.4\columnwidth]{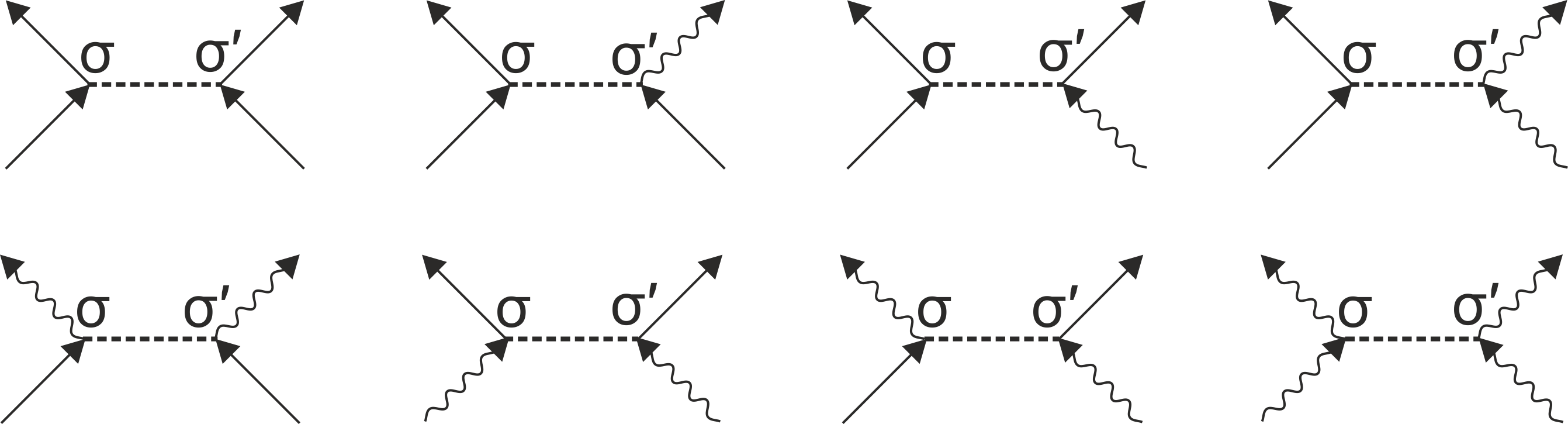}}
\caption{Possible types of the elementary graphs. Solid lines correspond to the bare Green's function $G^{(0)}$. Wavy lines describe emission and absorption of particles by the condensate. Dashed line stands for the interaction. The interaction conserves the spin of the particles, denoted by $\sigma$.}
\label{Elementars}
\end{figure*}

We consider a mixture of two bosonic species ($\sigma=\uparrow,\downarrow$) occupying the volume $V$ and characterized by the densities $n_\sigma=N_\sigma/V$ with $N_\sigma$ being the total number of particles in each component. As usual, we assume the thermodynamic limit $N_\sigma\rightarrow\infty$ and $V\rightarrow\infty$ with $n_\sigma$ being kept fixed. The second-quantized Hamiltonian of the system reads
\begin{widetext}
\begin{equation}
\label{Hamiltonian}
\begin{split}
\hat H&=\int\sum_\sigma\frac{\hbar^2}{2m_\sigma}\nabla\hat\Psi^\dagger_\sigma(\bm x)\nabla\hat\Psi_\sigma(\bm x) d\bm x+\frac{1}{2}\int\sum_{\sigma,\sigma'}\hat\Psi^{\dag}_{\sigma}(\bm x_1)\hat\Psi^{\dag}_{\sigma'}(\bm x_2)V_{\sigma\sigma'}(\bm x_1-\bm x_2)\hat\Psi_{\sigma}(\bm x_1)\hat\Psi_{\sigma'}(\bm x_2)d\bm x_1 d\bm x_2\\
&=\sum_{\textbf{p},\sigma}\frac{\hbar^2p^2}{2m_\sigma}\hat a_{\sigma, \textbf{p}}^{\dag} \hat a_{\sigma, \textbf{p}}+\frac{1}{2V}\sum_{\textbf p_1,\textbf p_2,\textbf{q},\sigma,\sigma^\prime}\hat a_{\sigma, \textbf p_1+\textbf q}^{\dag} \hat a_{\sigma^\prime,\textbf p_2-\textbf q}^{\dag} V_{\sigma\sigma^\prime}(\textbf{q})\hat a_{\sigma, \textbf p_1}\hat a_{\sigma^\prime,\textbf p_2}.
\end{split}
\end{equation}
\end{widetext}
Here $V_{\sigma\sigma'}(\bm x_1-\bm x_2)$ are the two-body interaction potentials with $\bm x$ being a $d$-dimensional coordinate and
\begin{equation}
V_{\sigma\sigma'}(\bm q)=\int e^{-i\bm q\bm x}V_{\sigma\sigma'}(\bm x)d\bm x
\end{equation}
are their Fourier transforms. The field operators $\hat\Psi_\sigma(\bm x)$ are related to the corresponding boson annihilation operators by
\begin{equation}\label{Psi1}
\hat\Psi_\sigma(\bm x)=\frac{1}{\sqrt{V}} \sum_{\textbf{p}} \hat a_{\sigma, \bm p} e^{i \bm p \bm x},
\end{equation}
and $\hat a_{\sigma, \bm p}$ obey
\begin{equation}
[\hat a_{\sigma, \bm p_1},\hat a_{\sigma', \bm p_2}^\dagger]=\delta_{\sigma\sigma',\bm p_1\bm p_2}.
\end{equation}             
With equal masses of different species 
\begin{equation}
\label{EqMasses}
m_\uparrow=m_\downarrow=m
\end{equation}
the model \eqref{Hamiltonian} has been applied to study resonant pairing of bright (dark) excitons in semiconductor heterostructures \cite{ResonantPairing} and formation of quantum droplets in a mixture of $\ket{m_F=-1}$ and $\ket{m_F=0}$ hyperfine states of the $F=1$ manifold of $^{39}$K \cite{Petrov2015, BoseMixture}. For the excitons one can additionally assume 
\begin{equation}
\label{SymmPotentials}
V_{\uparrow\uparrow}(\bm x)=V_{\downarrow\downarrow}(\bm x).
\end{equation}
Below we adopt the simplifying assumptions \eqref{EqMasses} and \eqref{SymmPotentials} in order to make our consideration more transparent. The general case of unequal masses and asymmetric pairwise potentials is discussed in the Appendix.

\section{General solution}

\subsection{Notations and the elementary graphs}

The arguments presented below are entirely based on the hypothesis of existence of a Bose-Einstein condensate in the ground state of the Hamiltonian \eqref{Hamiltonian}. This is usually justified \textit{a posteriori} for a weakly-interacting dilute system (the corresponding conditions will be presented in Section IV). In general, there is no condensate in 1D ($d=1$) even at the absolute zero temperature, and applicability of our results to this case should be discussed with care. We shall postpone such a discussion for future work.   

From a mathematical viewpoint, the presence of a condensate results in non-zero expectation values of the operators $\langle\hat \Psi_\sigma\rangle$. The condensate plays the role of a reservoir, which does not change its state upon increase or decrease of the number of particles $N_\sigma$ by one. As a consequence, the time evolution of the condensate wavefunctions is governed by the chemical potentials $\mu_\sigma$.   

Miscibility of the system means that both spin components occupy the same volume. The corresponding condition for a dilute gas is given by Eq. \eqref{miscibility} below. With the assumption \eqref{SymmPotentials} the equilibrium configuration corresponds to $n_\uparrow=n_\downarrow\equiv n$. The mixture thus can be characterized by a unique chemical potential $\mu$. 

The formalism of Green's functions in a Bose-Bose mixture can be developed along the lines of the spinless theory \cite{Beliaev}. We write the field operators in the form
\begin{equation}\label{Psi2}
\hat\Psi_\sigma(\bm x)=\hat\Psi_\sigma^\prime(\bm x) + \frac{\hat a_{0, \sigma}}{\sqrt{V}},
\end{equation}
where $\hat\Psi_\sigma^\prime$ stand for the non-condensed part and $\hat a_{0, \sigma}$ act on the macroscopically populated single-particle states with $\bm p=0$. The Green's functions are defined in terms of the non-condensed parts of the operators in the Heisenberg representation
\begin{equation}
\label{G}
G_{\sigma \sigma^\prime}(\mathsf x_1,\mathsf x_2) = - i \langle T \hat\Psi^\prime_\sigma(\mathsf x_1) \hat\Psi^{\prime \dag}_{\sigma^\prime}(\mathsf x_2) \rangle,
\end{equation}
where we have introduced the four-vectors $\mathsf x_i=(t_i, \bm x_i)$. For a uniform system one has
\begin{equation}
G_{\sigma \sigma^\prime}(\mathsf x_1,\mathsf x_2)=G_{\sigma \sigma^\prime}(\mathsf x),
\end{equation}
where $\mathsf x=\mathsf x_1-\mathsf x_2$. To describe absorption and emission of the particles by the condensate we shall also need the following auxiliary quantities
\begin{eqnarray}
 \label{F1}
 iF_{\sigma \sigma^\prime}(\mathsf x) &=& \langle N-2| T \hat\Psi^\prime_\sigma(\mathsf x_1) \hat\Psi^{\prime}_{\sigma^\prime}(\mathsf x_2) | N \rangle, \\ \label{F2}
 iF^\dag_{\sigma \sigma^\prime}(\mathsf x) &=& \langle N+2| T \hat\Psi^{\prime \dag}_\sigma(\mathsf x_1) \hat\Psi^{\prime \dag}_{\sigma^\prime}(\mathsf x_2) | N \rangle,
\end{eqnarray}
known as anomalous Green's functions \cite{Pitaevskii}. In what follows we shall use the momentum-space representation for the Green's function. The corresponding transformation is given by
\begin{equation}
G(\mathsf p)=\int e^{i \mathsf p \mathsf x}G(\mathsf x)d^4 \mathsf x,
\end{equation}
where $\mathsf p=(\omega, \bm p)$ and $\mathsf p\mathsf x=\omega t-\bm p\bm x$. It will also be convenient to use the modified Hamiltonian
\begin{equation}
\label{Hprime}
\hat H^\prime=\hat H-\mu \hat N
\end{equation}
in setting the time-dependence of the operators. For an ideal gas we obtain
\begin{equation}\label{G0}
  G^{(0)}_{\sigma \sigma^\prime} (\mathsf p) = \delta_{\sigma \sigma^\prime} G_{0}(\mathsf p)=\delta_{\sigma \sigma^\prime} \left[ \hbar\omega - \frac{\hbar^2p^2}{2m}+\mu+i0 \right]^{-1},
\end{equation}
where $\mu$ should be regarded as a free parameter.

Each diagram contributing to the expansion of $G_{\sigma\sigma^\prime}$ can be composed of the eight elementary graphs shown in Fig. (1). Wavy lines describe the emission and absorption of particles by the condensate. In calculations they are replaced by the factor $\sqrt{n_{\sigma,0}}$, where $n_{\sigma,0}$ is the $\sigma$-component of the condensate density. Dashed lines carry the factors $-iV_{\sigma\sigma'}(\bm q)$. Each vertex has a label $\sigma$ showing the spin of an incoming (outgoing) particle.

\begin{figure*}[t]
  \noindent\centering{
  \includegraphics[width=1.6\columnwidth] {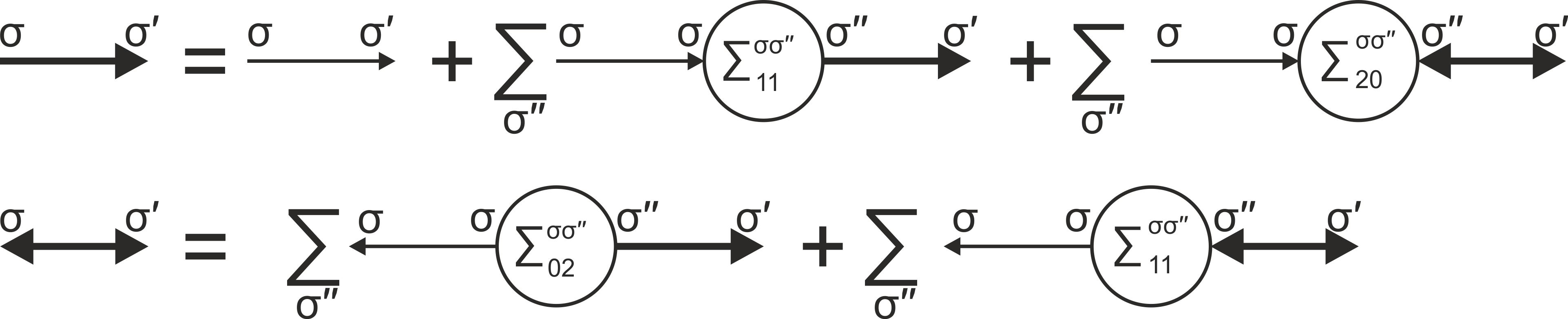}}
  \caption{Dyson equations. The Greens's functions (bold lines with arrows) couple to each other via the self-energies (circles). For each pair of spin indices $\sigma\sigma^\prime$ there are three types of self-energies characterized by different numbers of incoming (the left index in the lower row) and outgoing (the right index) lines.}
  \label{Dyson}
\end{figure*} 

\subsection{Dyson equations}

Though the interaction of two particles in a vacuum conserves the particle spin, the latter can be effectively changed after scattering off the condensate. As one can see from Fig. 1, already in the first order of perturbation theory there is a finite probability amplitude to find the particle in a state with a different $\sigma$. Formally, this results in appearance of the matrix elements $G_{\sigma\sigma^\prime}$ with $\sigma\neq\sigma^\prime$ for the Green's function of an interacting system. An accurate consideration of the higher order terms shows, that the resulting picture of many-body scattering processes can be recast in the graphical form shown in Fig. 2. The Greens's functions (bold lines with arrows) couple to each other via the self-energies (circles) obtained by summation of possible irreducible parts. There are three types of these parts for each pair of spin indices $\sigma\sigma^\prime$ differing by the number of incoming and outgoing continuous lines. By analogy with Ref. \cite{Beliaev}, we denote the resulting potentials by $\Sigma^{\sigma \sigma^{\prime}}_{11}$, $\Sigma^{\sigma \sigma^{\prime}}_{20}$ and $\Sigma^{\sigma \sigma^{\prime}}_{02}$. The graphical form in Fig. (2) then can be translated into the following system of Dyson equations
\begin{widetext}
\begin{subequations} \label{Dys}
\begin{align}
  G_{\sigma \sigma^\prime}(\mathsf p)=& G^{(0)}_{\sigma \sigma^\prime}(\mathsf p) + \sum_{\sigma^{\prime\prime}} G_{0}(\mathsf p) \Sigma^{\sigma \sigma^{\prime\prime}}_{11} G_{\sigma^{\prime\prime} \sigma^\prime}(\mathsf p) + \sum_{\sigma^{\prime\prime}} G_{0}(\mathsf p) \Sigma^{\sigma \sigma^{\prime\prime}}_{20} F^\dag_{\sigma^{\prime\prime} \sigma^\prime}(\mathsf p), \\ 
  F^\dag_{\sigma \sigma^\prime}(\mathsf p)=& \sum_{\sigma^{\prime\prime}} G_{0}(-\mathsf p) \Sigma^{\sigma \sigma^{\prime\prime}}_{02}(\mathsf p) G_{\sigma^{\prime\prime} \sigma^\prime}(\mathsf p) + \sum_{\sigma^{\prime\prime}} G_{0}(-\mathsf p) \Sigma^{\sigma \sigma^{\prime\prime}}_{11}(-\mathsf p) F^\dag_{\sigma^{\prime\prime} \sigma^\prime}(\mathsf p).
  \end{align}
\end{subequations}
\end{widetext}
By noticing that the equations with different $\sigma^\prime$ decouple from each other, we can write the system \eqref{Dys} in the useful form
\begin{widetext}
\begin{equation}\label{Matr1}
  \left[
    \begin{array}{cccc}
      G^{-1}_0(\mathsf p) - \Sigma^{\uparrow \uparrow}_{11}(\mathsf p) & - \Sigma^{\uparrow \downarrow}_{11}(\mathsf p) & - \Sigma^{\uparrow \uparrow}_{20}(\mathsf p) & - \Sigma^{\uparrow \downarrow}_{20}(\mathsf p) \\
      - \Sigma^{\downarrow \uparrow}_{11}(\mathsf p) & G^{-1}_0(\mathsf p) - \Sigma^{\downarrow \downarrow}_{11}(\mathsf p) & - \Sigma^{\downarrow \uparrow}_{20}(\mathsf p) & - \Sigma^{\downarrow \downarrow}_{20}(\mathsf p) \\
      - \Sigma^{\uparrow \uparrow}_{02}(\mathsf p) & - \Sigma^{\uparrow \downarrow}_{02}(\mathsf p) & G^{-1}_0(-\mathsf p) - \Sigma^{\uparrow \uparrow}_{11}(-\mathsf p) & - \Sigma^{\uparrow \downarrow}_{11}(-\mathsf p) \\
      - \Sigma^{\downarrow \uparrow}_{02}(\mathsf p) & - \Sigma^{\downarrow \downarrow}_{02}(\mathsf p) & - \Sigma^{\downarrow \uparrow}_{11}(-\mathsf p) & G^{-1}_0(-\mathsf p) - \Sigma^{\downarrow \downarrow}_{11}(-\mathsf p) \\
    \end{array}
  \right] \left[
            \begin{array}{c}
              G_{\uparrow \uparrow}(\mathsf p) \\
              G_{\downarrow \uparrow}(\mathsf p) \\
              F^\dag_{\uparrow \uparrow}(\mathsf p) \\
              F^\dag_{\downarrow \uparrow}(\mathsf p) \\
            \end{array}
          \right]
          = \left[
              \begin{array}{c}
                1 \\
                0 \\
                0 \\
                0 \\
              \end{array}
            \right].
\end{equation}
\end{widetext}

\begin{figure*}[t]
  \noindent\centering{
  \includegraphics[width=1.2\columnwidth] {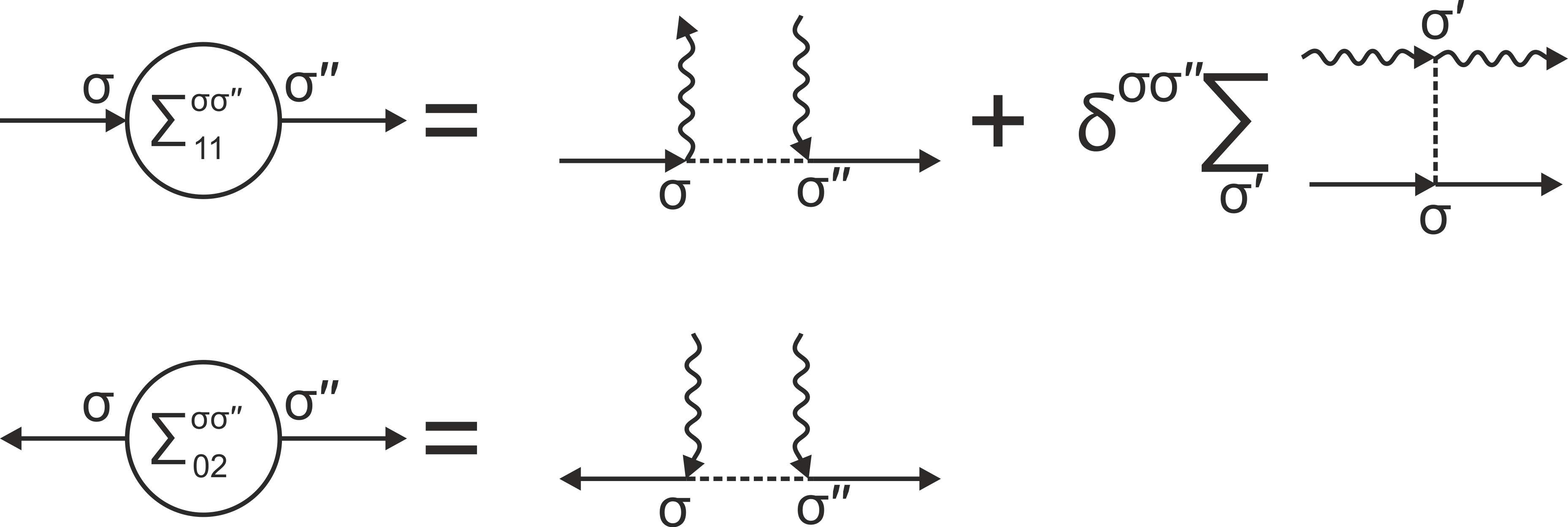}}
  \caption{First-order diagrams in the expansion of the self-energies $\Sigma^{\sigma\sigma^\prime}_{11}(\mathsf p)$ and $\Sigma^{\sigma\sigma^\prime}_{02}(\mathsf p)$, defining the chemical potential and the spectrum of elementary excitations according to Eq. \eqref{Mu1} and Eq. \eqref{Spec1}, respectively.}
  \label{Firstorder}
\end{figure*} 

Furthermore, by virtue of \eqref{SymmPotentials}, one has $\Sigma^{\uparrow \uparrow}_{11}(\mathsf p)=\Sigma^{\downarrow \downarrow}_{11}(\mathsf p)$. Note also that $\Sigma^{\sigma \sigma^{\prime\prime}}_{02}(\mathsf p)= \Sigma^{\sigma \sigma^{\prime\prime}}_{20}(\mathsf p)$, because the relevant diagrams differ only by the direction of the wavy lines. This allows us to write Eq. \eqref{Matr1} in the spinor form
\begin{widetext}
\begin{equation}\label{Matr2}
  \left[
    \begin{array}{cc}
      G_1^{-1}(\mathsf p) \hat\sigma_0 - \Sigma^{\uparrow \downarrow}_{11}(\mathsf p) \hat\sigma_1 & - \Sigma^{\uparrow \uparrow}_{20}(\mathsf p) \hat\sigma_0 - \Sigma^{\uparrow \downarrow}_{20}(\mathsf p) \hat\sigma_1 \\
      - \Sigma^{\uparrow \uparrow}_{20}(\mathsf p) \hat\sigma_0 - \Sigma^{\uparrow \downarrow}_{20}(\mathsf p) \hat\sigma_1 & G^1_{-1}(-\mathsf p) \hat\sigma_0 - \Sigma^{\uparrow \downarrow}_{11}(-\mathsf p) \hat\sigma_1 \\
    \end{array}
  \right] \left[
            \begin{array}{c}
              \varphi \\
              \chi \\
            \end{array}
          \right] = \left[
            \begin{array}{c}
              \alpha \\
              0 \\
            \end{array}
          \right],
\end{equation}
\end{widetext}
where
\begin{equation}\label{G11}
  G^{-1}_1(\mathsf p)=G^{-1}_0(\mathsf p) - \Sigma^{\uparrow \uparrow}_{11}(\mathsf p)
\end{equation}
and
\begin{equation}\label{Pauli1}
  \hat\sigma_0=\left[
             \begin{array}{cc}
               1 & 0 \\
               0 & 1 \\
             \end{array}
           \right], \quad \hat\sigma_1=\left[
             \begin{array}{cc}
               0 & 1 \\
               1 & 0 \\
             \end{array}
           \right],\quad\alpha=\left[
           \begin{array}{c}
           1\\
           0\\
           \end{array}
           \right].
\end{equation}

The system \eqref{Matr2} then can be solved by using the identity
\begin{equation*}
a^2-b^2=(a \hat\sigma_0 - b \hat\sigma_1)(a \hat\sigma_0 + b \hat\sigma_1).
\end{equation*}
We first use the second row in \eqref{Matr2} to express $\chi$ via $\varphi$, and then substitute it into the first row. We find that all Green's functions have the denominator $D_1(\mathsf p)D_2(\mathsf p)$, where
\begin{widetext}
\begin{eqnarray}
\label{D1P}
  D_1(\mathsf p) &=& (G^{-1}_1(\mathsf p)-\Sigma^{\uparrow \downarrow}_{11}(\mathsf p))(G^{-1}_1(-\mathsf p)-\Sigma^{\uparrow \downarrow}_{11}(-\mathsf p))-(\Sigma^{\uparrow \uparrow}_{20}(\mathsf p)+ \Sigma^{\uparrow \downarrow}_{20}(\mathsf p))^2, \\ \label{D2P}
  D_2(\mathsf p) &=& (G^{-1}_1(\mathsf p)+\Sigma^{\uparrow \downarrow}_{11}(\mathsf p))(G^{-1}_1(-\mathsf p)+\Sigma^{\uparrow \downarrow}_{11}(-\mathsf p))-(\Sigma^{\uparrow \uparrow}_{20}(\mathsf p)- \Sigma^{\uparrow \downarrow}_{20}(\mathsf p))^2.
\end{eqnarray}
\end{widetext}
In terms of these quantities the solution of Eq.\eqref{Matr1} can be written as
\begin{widetext}
\begin{subequations}
\label{GF}
\begin{align}
 \label{Gud}
  G_{\sigma\sigma^\prime}= & \frac{1}{2} \left( \frac{G^{-1}_1(-\mathsf p)-\Sigma^{\uparrow \downarrow}_{11}(-\mathsf p)}{D_1(\mathsf p)} \pm \frac{G^{-1}_1(-\mathsf p)+\Sigma^{\uparrow \downarrow}_{11}(-\mathsf p)}{D_2(\mathsf p)} \right),\\ \label{Fud}
  F^\dag_{\sigma\sigma^\prime}=& \frac{1}{2} \left( \frac{\Sigma^{\uparrow \uparrow}_{20}(\mathsf p)+ \Sigma^{\uparrow \downarrow}_{20}(\mathsf p)}{D_1(\mathsf p)} \pm \frac{\Sigma^{\uparrow \uparrow}_{20}(\mathsf p)- \Sigma^{\uparrow \downarrow}_{20}(\mathsf p)}{D_2(\mathsf p)} \right),
\end{align}
\end{subequations}
\end{widetext}
where ``$+$'' should be used for $\sigma=\sigma^\prime$ and ``$-$'' for $\sigma \neq \sigma^\prime$.

With the result \eqref{GF} one can readily express the chemical potential of the system in terms of the self-energies. We notice, that in the long-wavelength limit the above-condensate part of the field operator can be written as $\hat\Psi^\prime_\sigma \approx i \sqrt{n_{0\sigma}} \hat\Phi_\sigma$, where the operator $\hat\Phi_\sigma$ is the phase of the condensate. Hence, one has $F^\dag_{\uparrow \uparrow} \approx -G_{\uparrow\uparrow}$. On the other hand, as a consequence of the symmetry breaking we may expect two gapless Goldstone modes in the elementary excitation spectrum, which implies the condition $D_1(0)D_2(0)=0$. By noticing also that $\Sigma^{\uparrow \downarrow}_{20}(0)= \Sigma^{\uparrow \downarrow}_{11}(0)$, we obtain
\begin{equation}\label{Mu1}
  \mu=\Sigma^{\uparrow \uparrow}_{11}(0) - \Sigma^{\uparrow \uparrow}_{20}(0).
\end{equation}
The result \eqref{Mu1} provides the dependence of the chemical potential on the density $n_0$ of the condensate components, and together with the well-known formula \cite{AGD}
\begin{equation}
\label{depletion}
n=n_0+\frac{i}{(2\pi)^{d+1}}\lim_{{t}\rightarrow{-0}}\int G_{\uparrow\uparrow}(\mathsf p)e^{-i\omega t}d\mathsf p
\end{equation}
allows one to calculate $\mu$ as a function of the total density $n$, which includes the above-condensate particles.

\subsection{Elementary excitation spectrum}

According to the general theorem \cite{AGD} the spectrum of elementary excitations of the system can be obtained from the poles of the Green's functions. By solving $D_1(\bm p,\omega)D_2(\bm p,\omega)=0$ with respect to $\omega$ we find two branches for the excitations of the particle type (we shall omit the hole excitations for brevity):
\begin{widetext}
\begin{equation}
\label{Spec1}
\hbar\omega(\bm p)=\sqrt{(\hbar^2 p^2/2m+\Sigma_s^{\uparrow\uparrow}(\mathsf p)\pm\Sigma_s^{\uparrow\downarrow}(\mathsf p)-\mu)^2-(\Sigma^{\uparrow \uparrow}_{20}(\mathsf p) \pm \Sigma^{\uparrow \downarrow}_{20}(\mathsf p))^2}+\Sigma_a^{\uparrow\uparrow}(\mathsf p)\pm\Sigma_a^{\uparrow\downarrow}(\mathsf p)
\end{equation}
\end{widetext}
where we have introduced
\begin{equation}
\Sigma_{s,a}^{\sigma\sigma^\prime}(\mathsf p)=\frac{\Sigma^{\sigma\sigma^\prime}_{11}(\mathsf p)\pm\Sigma^{\sigma\sigma^\prime}_{11}(-\mathsf p)}{2}.
\end{equation} 
Strictly speaking, Eq. \eqref{Spec1} is a transcendental equation on $\omega$. As we shall see, to a good accuracy one can neglect the dependence of the self-energies on $\omega$ in the dilute regime. Thus, in 3D it is common to model the system by a hypothetical weakly-interacting gas characterized by $V_{\sigma\sigma^\prime}(\bm q)=g_{\sigma\sigma^\prime}$ for $q R_e\ll 1$ with $R_e$ being the interaction radius. One can then approximate the self-energies by few first-order diagrams shown in Fig. \ref{Firstorder}. We find $\Sigma^{\uparrow \uparrow}_{11}=n(2g_{\uparrow \uparrow}+g_{\uparrow \downarrow})$, $\Sigma^{\uparrow \downarrow}_{11}=n g_{\uparrow \downarrow}$, $\Sigma^{\uparrow \uparrow}_{20}= n g_{\uparrow \uparrow}$, $\Sigma^{\uparrow \downarrow}_{20}= n g_{\uparrow \downarrow}$, which, upon substitution into \eqref{Spec1} yields the well-known result
\begin{subequations}\label{SpecB1}
\begin{align}
\hbar\omega(\bm p)&=\sqrt{\left( \frac{\hbar^2 p^2}{2m} \right)^2 + \frac{\hbar^2p^2}{m}n(g_{\uparrow\uparrow}\pm g_{\uparrow\downarrow})}\\
\mu&=n(g_{\uparrow \uparrow}+g_{\uparrow \downarrow})
\end{align}
\end{subequations}
for the spectrum and the chemical potential of the system. Relation of the constants $g_{\uparrow \uparrow}$ and $g_{\uparrow \downarrow}$ to the characteristics of the original model will be discussed below.

To conclude this part, let us point out an important symmetry property of the formula \eqref{Spec1}. In the long-wavelength limit $\mathsf p\rightarrow 0$ we can use the Gavoret-Nozieres type of arguments \cite{Gavoret} to obtain the following relations
\begin{widetext}
\begin{subequations}
\label{GNrelations}
\begin{align}
\Sigma_s^{\uparrow\uparrow}(\mathsf p)-\Sigma_s^{\uparrow\downarrow}(\mathsf p)-\Sigma^{\uparrow \uparrow}_{20}(\mathsf p)+\Sigma^{\uparrow \downarrow}_{20}(\mathsf p)-\mu&=\frac{\hbar^2 p^2}{2m}\left(\frac{n^\prime}{n_0}-\frac{\rho_{\uparrow\downarrow}}{mn_0}\right)\\
\Sigma_s^{\uparrow\uparrow}(\mathsf p)-\Sigma_s^{\uparrow\downarrow}(\mathsf p)+\Sigma^{\uparrow \uparrow}_{20}(\mathsf p)-\Sigma^{\uparrow \downarrow}_{20}(\mathsf p)-\mu&=2(\Sigma^{\uparrow \uparrow}_{20}(\mathsf p)-\Sigma^{\uparrow \downarrow}_{20}(\mathsf p))+\frac{\hbar^2 p^2}{2m}\left(\frac{n^\prime}{n_0}-\frac{\rho_{\uparrow\downarrow}}{mn_0}\right),
\end{align}
\end{subequations}
\end{widetext}
where $n^\prime=n-n_0$ is the quantum depletion of the condensate and $\rho_{\uparrow\downarrow}$ is the so-called \textit{superfluid drag} due to Andreev-Bashkin effect \cite{AB, ABnote}. For spin-independent interactions one has $\Sigma^{\uparrow \uparrow}_{20}(\mathsf p)=\Sigma^{\uparrow \downarrow}_{20}(\mathsf p)$ and $\Sigma_a^{\uparrow\uparrow}(\mathsf p)=\Sigma_a^{\uparrow\downarrow}(\mathsf p)$, and, by virtue of \eqref{GNrelations}, the lower branch in Eq. \eqref{Spec1} takes the form
\begin{equation}
\label{magnon}
\hbar\omega_\mathrm{m} (\bm p)=\frac{\hbar^2 p^2}{2m_\ast},
\end{equation}
where
\begin{equation}
\label{magnonmass}
m_\ast=\frac{n_0}{n}\frac{m}{1-\rho_{\uparrow\downarrow}/mn}
\end{equation}
is the effective mass. 

An energy spectrum quadratic in $\bm k$ is what one would expect on general grounds for an arbitrary multicomponent superfluid \cite{Halperin1}. The dispersion of the type \eqref{magnon} describes the excitations analogous to the spin waves in a Heisenberg ferromagnet \cite{Halperin2}.

In the asymmetric case $\Sigma^{\uparrow \uparrow}_{20}(\mathsf p)>\Sigma^{\uparrow \downarrow}_{20}(\mathsf p)$ one finds $\hbar\omega_\mathrm{m} (\bm p)=\hbar c_\mathrm{m} p$ with
\begin{equation}
\label{spinvelocity}
c_\mathrm{m}^2=\frac{(nm-\rho_{\uparrow\downarrow})}{m^2}\frac{[\Sigma^{\uparrow \uparrow}_{20}(\mathsf p)-\Sigma^{\uparrow \downarrow}_{20}(\mathsf p)]}{n_0}
\end{equation}
being the spin wave velocity. For the model potential $V_{\sigma\sigma^\prime}(\bm q)=g_{\sigma\sigma^\prime}$ considered above the result \eqref{spinvelocity} matches the hydrodynamic formula of Ref. \cite{Nespolo}. One can see that the entrainment slows the propagation of magnons.     

\begin{figure*}[t]
  \noindent\centering{
  \includegraphics[width=1.2\columnwidth] {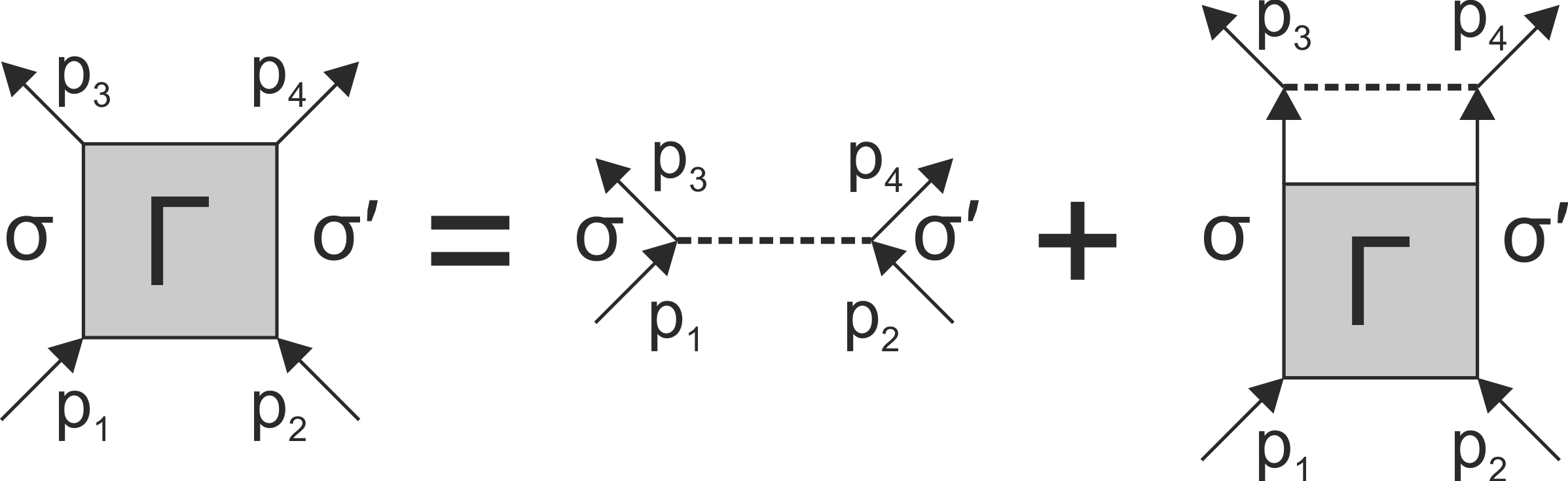}}
  \caption{Graphical equation for the effective inetraction in the dilute regime.}
  \label{Vertex}
\end{figure*}

\section{Dilute regime}

By analogy with the spinless theory \cite{Beliaev, Lozovik}, estimation of the integrals over the internal momenta in the graphs for $\Sigma$'s shows, that to the lowest order in $\beta$ only the ladders should be retained. These obey the diagrammatic rule shown schematically in Fig. 4. One can readily recognise the structure typical for the scattering problem of two particles in vacuum. Indeed, by introducing the relative $\mathsf p_1-\mathsf p_2=2\mathsf k$, $\mathsf p_3-\mathsf p_4=2\mathsf k^\prime$ and total $\mathsf p_1+\mathsf p_2=\mathsf p_3+\mathsf p_4=\mathsf P=(\Omega, \bm{P})$ momenta and taking advantage of the fact that $V_{\sigma\sigma^\prime}(\bm q)$ does not depend on frequency, one can recast Fig. 4 in the form
\begin{multline}\label{Vert2}
T_{\sigma\sigma^\prime}(\bm k^\prime,\bm k; z) = \frac{1}{(2\pi)^d}V_{\sigma\sigma^\prime}(\textbf{k}^\prime-\bm k)\\
 +\frac{1}{(2\pi)^d} \int \frac{V_{\sigma\sigma^\prime}(\textbf{k}^\prime - \textbf{k}^{\prime\prime})}{z-E_{\bm k^{\prime\prime}}}T_{\sigma\sigma^\prime}(\bm k^{\prime\prime},\bm k; z) d\bm k^{\prime\prime},
\end{multline}
where $E_{\bm k^{\prime\prime}}=\hbar^2 k^{\prime\prime 2}/m$ and
\begin{equation}\label{Kappa1}
z=\Omega - \frac{P^2}{4m} + 2\mu+i0.
\end{equation}
This allows one to identify the quantity
\begin{equation*}
 T_{\sigma\sigma^\prime}(\bm k^\prime,\bm k; z)\equiv\frac{1}{(2\pi)^d}\Gamma(\mathsf p_1,\mathsf p_2;\mathsf p_3,\mathsf p_4)
 \end{equation*}
 with the matrix elements of the $T_{\sigma\sigma^\prime}$-operator of the quantum scattering theory \cite{Taylor}. Furthermore, the $T_{\sigma\sigma^\prime}$-operator can be expressed in terms of the off-shell scattering amplitude defined by
\begin{equation}
f_{\sigma\sigma^\prime}(\bm k^\prime,\bm k)=-(2\pi)^2\frac{m}{2\hbar^2}T_{\sigma\sigma^\prime}(\bm k^\prime,\bm k; E_{\bm k}+i0).
\end{equation}
The corresponding relation reads
\begin{widetext}
\begin{equation}
\label{Vert3}
T_{\sigma\sigma^\prime}(\bm k^\prime,\bm k; z)=-\frac{1}{(2\pi)^2}\frac{2\hbar^2}{m}\Bigl [f_{\sigma\sigma^\prime}^{\ast}(\bm k,\bm k^\prime)-
\frac{1}{(2\pi)^2}\frac{2\hbar^2}{m}\int f_{\sigma\sigma^\prime}(\bm k^\prime,\bm q)f_{\sigma\sigma^\prime}^{\ast}(\bm k,\bm q)\left (\frac{1}{E_{\bm q}-E_{\bm k^\prime}+i0}+\frac{1}{z-E_{\bm q}}\right)d\bm q\Bigr].
\end{equation}
\end{widetext}  
The self-energies are defined by the special matrix elements of the $T$-operator obtained by letting two out of the four particles belong to the condensate:
\begin{equation}
\label{SigmaViaTs}
\begin{split}
\Sigma^{\sigma\sigma^\prime}_{11}(\pm\mathsf p)&=(2\pi)^d n_0 [T_{\sigma\sigma^\prime}(\mp\bm p/2,\pm\bm p/2;z_\pm)\\
&+\delta_{\sigma\sigma^\prime}\sum_{\sigma^{\prime\prime}}T_{\sigma\sigma^{\prime\prime}}(\pm\bm p/2,\pm\bm p/2;z_\pm)]\\
\Sigma^{\sigma\sigma^\prime}_{20}(\mathsf p)&=(2\pi)^d n_0 T_{\sigma\sigma^\prime}(0,\bm p;2\mu+i0),
\end{split}
\end{equation}
where
\begin{equation}
z_\pm=\pm\hbar\omega-
\frac{\hbar^2 p^2}{4m}+2\mu+ i0.
\end{equation}
The chemical potential satisfies the transcendental equation
\begin{equation}
\label{mu}
\mu=(2\pi)^d n_0[T_{\uparrow\uparrow}(0,0; 2\mu+i0)+T_{\uparrow\downarrow}(0,0; 2\mu+i0)].
\end{equation}
Assuming slow dependence of  $T_{\sigma\sigma^\prime}$ on $\mu$ and $n_0\approx n$ one can write
\begin{equation}
E_{\mathrm{mix}}=\int \mu dN=\frac{(2\pi)^dN^2 (T_{\uparrow\uparrow}+T_{\uparrow\downarrow})}{4V},
\end{equation}
where we have used the shortcut $T_{\uparrow\uparrow}\equiv T_{\uparrow\uparrow}(0,0; 2\mu+i0)$. On the other hand, for a phase separated configuration one has
\begin{equation}
E_{\mathrm{separ}}=\frac{(2\pi)^dN^2 T_{\uparrow\uparrow}}{2V}.
\end{equation}
Comparing the two energies we find $T_{\uparrow\downarrow}<T_{\uparrow\uparrow}$ as the condition of miscibility. More generally,
\begin{equation}
\label{miscibility}
T_{\uparrow\downarrow}^2<T_{\uparrow\uparrow}T_{\downarrow\downarrow},
\end{equation}
which applies also to the spin-imbalanced configurations $n_\uparrow\neq n_\downarrow$ (see Appendix B). Further conclusions depend on the dimensionality of the problem.

\subsection{3D gas}

In the 3D geometry to the first order in $\beta$ one can neglect the integral term in Eq. \eqref{Vert3}. Taking into account the invariance of the on-shell scattering amplitude with respect to the time reversal, we obtain
\begin{equation}
\label{Sigmas3D}
\begin{split}
\Sigma_{a}^{\sigma\sigma^\prime}(\mathsf p)&=0\\
\Sigma_s^{\uparrow\uparrow}(\mathsf p)\pm\Sigma_s^{\uparrow\downarrow}(\mathsf p)&=-\frac{8\pi\hbar^2 n_0}{m}[f_{\uparrow\uparrow}^{+}(\bm p/2,\bm p/2)+f_{\uparrow\downarrow}^{\pm}(\bm p/2,\bm p/2)]\\
\Sigma^{\sigma\sigma^\prime}_{20}(\mathsf p)&=-\frac{4\pi\hbar^2 n_0}{m}f_{\sigma\sigma^\prime}(0,\bm p)\\
\mu&=-\frac{4\pi\hbar^2 n_0}{m}[f_{\uparrow\uparrow}(0,0)+f_{\uparrow\downarrow}(0,0)],
\end{split}
\end{equation}
where we have defined
\begin{equation}
f_{\sigma\sigma^\prime}^{\pm}(\bm k^\prime,\bm k)=\frac{1}{2}[f_{\sigma\sigma^\prime}(\bm k^\prime,\bm k)\pm f_{\sigma\sigma^\prime}(-\bm k^\prime,\bm k)].
\end{equation}
At small momenta the leading contribution to the scattering is in the $s$-wave scattering channel, and the $s$-wave scattering amplitude is known to approach the constant value \cite{Taylor}
\begin{equation}
\label{f3D}
f_{\sigma\sigma'}(\bm k^\prime,\bm k)=-a_{\sigma\sigma'},
\end{equation}
known as the $s$-wave scattering length. Substitution of \eqref{f3D} into \eqref{Sigmas3D} yields the elementary excitation spectrum and the chemical potential of the type \eqref{SpecB1} with 
\begin{equation}
g_{\sigma\sigma^\prime}=\frac{4\pi\hbar^2 a_{\sigma\sigma^\prime}}{m}.
\end{equation}

The same result can be obtained by solving linearized equations of motion for the small-amplitude oscillations of the classical fields $\Psi_\sigma$ obtained from the Hamiltonian \eqref{Hamiltonian} where one substitutes $g_{\sigma\sigma^\prime}$ in lieu of  $V_{\sigma\sigma^\prime}(\bm q)$. Such treatment of the low-energy excitations is quite common \cite{Petrov2015, Goldstein, Berman} and is sometimes extended to momentum-dependent phenomenological potentials $g_{\sigma\sigma^\prime}(\bm q)$ as well \cite{ResonantPairing, Alexandrov, Eckardt}. Below we present the result of our theory which escapes this simplified approach.

Consider again the lower branch of the spectrum \eqref{Spec1} and assume the interaction potential to be not dependent on the particle's spin, so that $f_{\uparrow\uparrow}(\bm k^\prime,\bm k)=f_{\uparrow\downarrow}(\bm k^\prime,\bm k)\equiv f(\bm k^\prime,\bm k)$. By using the relations \eqref{Sigmas3D} we obtain
\begin{equation}
\label{DiluteMagnon}
\hbar\omega_\mathrm{m}(\bm p)=\frac{\hbar^2 p^2}{2m}-\frac{8\pi\hbar^2 n_0}{m}[f(\bm p/2,\bm p/2)-f(0,0)].
\end{equation}
The second term in the above equation for the magnon dispersion does not appear if one uses a standard hydrodynamic approach. Indeed, mere Fourier-expansion of the small-amplitude oscillations of the order parameter would yield the equation having the structure of \eqref{SpecB1}. For identical inter- and intra-species interactions the density-dependent term vanishes and one gets $\hbar\omega_{\mathrm m}(\bm p)\equiv \hbar^2 p^2/2m$. In the weakly-interacting limit $n a^3\ll 1$, the result \eqref{DiluteMagnon} can be reproduced if instead one applies the canonical Bogoliubov transformation to an \textit{ersatz} Hamiltonian (see Appendix A)
\begin{widetext}
\begin{equation}
\label{DiluteHamiltonian}
\hat H_\ast=\sum_{\textbf{p},\sigma}\frac{\hbar^2 p^2}{2m}\hat a_{\sigma, \textbf{p}}^{\dag} \hat a_{\sigma, \textbf{p}}+\frac{1}{2V}\sum_{\textbf k,\textbf p,\textbf q,\sigma,\sigma^\prime}\hat a_{\sigma, \textbf k+\textbf p}^{\dag} \hat a_{\sigma^\prime,\textbf k-\textbf p}^{\dag} g_{\sigma\sigma^\prime}(\bm p, \bm q)\hat a_{\sigma, \bm k+\bm q}\hat a_{\sigma^\prime,\bm k-\bm q},
\end{equation}
\end{widetext}
where
\begin{equation}
g_{\sigma\sigma^\prime}(\bm p, \bm q)\equiv-\frac{4\pi\hbar^2}{m} f_{\sigma\sigma^\prime}(\bm p, \bm q)
\end{equation}
are the properly defined effective potentials.
\begin{figure*}[t]
  \noindent\centering{
  \includegraphics[width=1.2\columnwidth] {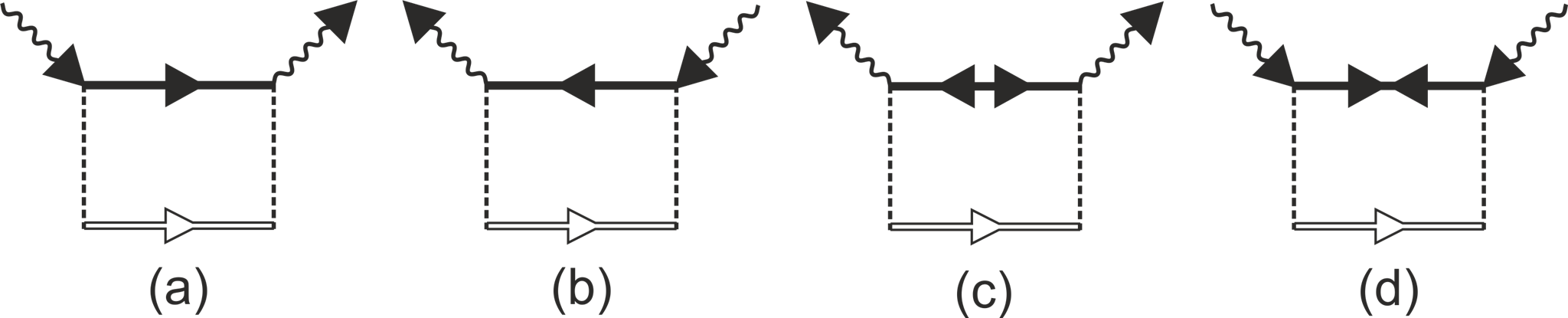}}
  \caption{Second-order graphs for the magnon self-energy due to interaction with phonons. Bold black and empty lines are used for the phonon \eqref{PhononG} and magnon \eqref{MagnonG} Green's functions, respectively. Wavy lines carry the factor $\sqrt{n_0}$. The picture is fully analogous to the second-order perturbative treatment of an impurity in a one-component Bose-Einstein condensate (Bose polaron) done in \cite{Christensen}.}    
  \label{Polaron}
\end{figure*}    
  
It would be wrong to identify the low-momentum expansion of the tail in Eq. \eqref{DiluteMagnon} with the drag density, as this expansion yields a subleading order with respect to the quantum depletion $n'\sim \sqrt{na^3}$ which enters the formula \eqref{magnonmass}. The leading correction to the magnon mass comes from the second-order approximation in $\beta$. For a weakly-interacting gas the result can be obtained by considering interaction of magnons with the Bogoliubov phonon modes. The bare propagators in this picture take the form
\begin{equation*}
\begin{split}
G_{\sigma\sigma'}(\mathsf p)&=\frac{1}{2}[G_\mathrm{ph}(\mathsf p)\pm G_\mathrm{m}(\mathsf p)]\\
F_{\sigma\sigma'}^\dagger(\mathsf p)&=\frac{1}{2}F_\mathrm{ph}^\dagger(\mathsf p),
\end{split}
\end{equation*}
with
\begin{equation}
\label{PhononG}
\begin{split}
G_\mathrm{ph}(\mathsf p)&=\frac{u_{\bm p}^2}{\hbar\omega-\hbar\omega_\mathrm{ph}(\bm p)+i0}-\frac{\upsilon_{\bm p}^2}{\hbar\omega+\hbar\omega_\mathrm{ph}(\bm p)-i0}\\
F_\mathrm{ph}^\dagger(\mathsf p)&=\frac{u_{\bm p} \upsilon_{\bm p}}{(\hbar\omega-\hbar\omega_\mathrm{ph}(\bm p)+i0)(\hbar\omega+\hbar\omega_\mathrm{ph}(\bm p)-i0)}
\end{split}
\end{equation}
and
\begin{equation}
\label{MagnonG}
G_\mathrm{m}(\mathsf p)=\frac{1}{\hbar\omega-\hbar^2 p^2/2m}
\end{equation}
being the phonon and the magnon Green's functions, respectively. Here
\begin{equation}
\label{BogoliubovCoeff}
\begin{split}
u_{\bm p}&=\sqrt{\frac{\hbar^2 p^2/2m+2ng}{\hbar\omega_\mathrm{ph}(\bm p)}+1}\\
\upsilon_{\bm p}&=-\sqrt{\frac{\hbar^2 p^2/2m+2ng}{\hbar\omega_\mathrm{ph}(\bm p)}-1}
\end{split}
\end{equation}
are the Bogoliubov coefficients for the phonon part. At this level of approximation we neglect the dependence of the effective potentials on the momenta [Eq. \eqref{f3D} for the scattering lengths with $a_{\uparrow\uparrow}=a_{\uparrow\downarrow}\equiv a$ and $g\equiv 4\pi\hbar^2 a/m$]  and take $n=n_0$. Retaining the terms cubic and quartic in $\hat a_{\sigma,\bm p}$ with $\bm p\neq 0$ in the Hamiltonian \eqref{DiluteHamiltonian} and substituting 
\begin{equation}
\label{ReducedBogoliubov}
\begin{split}
\hat a_{\uparrow,\bm p}&=\frac{1}{\sqrt{2}}(u_{\bm p} \hat b_{\bm p}+\upsilon_{\bm p} \hat b_{-\bm p}^\dagger+\hat c_{\bm p})\\
\hat a_{\downarrow,\bm p}&=\frac{1}{\sqrt{2}}(u_{\bm p} \hat b_{\bm p}+\upsilon_{\bm p} \hat b_{-\bm p}^\dagger-\hat c_{\bm p}),
\end{split}
\end{equation}
we get the magnon Hamiltonian
\begin{equation}
\label{MagnonHamiltonian}
\begin{split}
\hat H_\mathrm{m}&=\sum_{\textbf{p}}\frac{\hbar^2 p^2}{2m}\hat c_{ \textbf{p}}^{\dag} \hat c_{\textbf{p}}+\\
\frac{g}{V}&\sum_{\textbf k,\textbf p,\textbf q}\hat c_{\textbf k+\textbf p}^{\dag} \hat c_{\textbf k-\textbf p}^{\dag} \hat c_{\bm k+\bm q}\hat c_{\bm k-\bm q}+\hat H_{\mathrm{m}-\mathrm{ph}},
\end{split}
\end{equation}
where the last term
\begin{equation}
\label{PhononMagnon}
\hat H_{\mathrm{m}-\mathrm{ph}}=\frac{g}{V}\sqrt{\frac{N}{2}}\sum_{\textbf p,\textbf q}[\hat c_{\textbf p+\textbf q}^{\dag}\hat c_{\bm q}(u_p \hat b_{\bm p}+\upsilon_{\bm p} \hat b_{-\bm p}^\dagger)+h.c.]
\end{equation} 
describes the interaction of magnons with phonons. 

At zero temperature the second term in \eqref{MagnonHamiltonian} does not yield renormalization of the magnon mass. We are thus left with the second-order contribution of \eqref{PhononMagnon} shown in Fig. \ref{Polaron}. We notice, that the same graphs appear in the perturbation theory of a mobile impurity in a single-component condensate (Bose polaron) \cite{Christensen}. The magnon now drags a cloud of phonons, which increases its effective mass. According to the general formula \eqref{magnonmass} the change in the mass is directly related to the Andreev-Bashkin entrainment effect. Evaluation of the graphs in Fig. \ref{Polaron} yields
\begin{equation}
\label{DiluteDrag}
\frac{\rho_{\uparrow\downarrow}}{mn}=\sqrt{\frac{2}{\pi}}\frac{64}{45}\sqrt{na^3}.
\end{equation}
The same formula for the superfluid drag density has been obtained in the earlier works \cite{Pastukhov1, Fil} by using hydrodynamic approaches. Hence, our consideration establishes a link between the effect of entrainment and the physics of Bose polarons.
 
Let us now follow Beliaev \cite{Beliaev} in considering the behaviour of the magnon dispersion \eqref{DiluteMagnon} in the high-momentum region $pa\sim 1$. By using the well-known result 
\begin{equation}
\label{f3Dk}
f_0(k)=-\frac{\sin (k a)}{k} e^{-i k a}
\end{equation} 
for the $s$-wave part of $f(\bm k^\prime,\bm k)$ at the mass shell we get
\begin{equation}
\label{MagnonHighMomenta}
\hbar\omega_\mathrm{m}(\bm p)=\frac{\hbar^2 p^2}{2m}+\frac{8\pi\hbar^2 n_0 a}{m}\left[\frac{\sin (pa)}{pa} -1\right],
\end{equation}
where we have omitted the imaginary part describing the damping of quasiparticles. Very similar expression can be obtained for the phonon (upper) branch of the dispersion. In that latter case Beliaev noticed, that if one formally allows the parameter $na^3$ to approach the unity, the spectrum develops a roton minimum. Such hypothetical state would mimic the superfluid Helium, rotonization of the spectrum being a signature of strong correlations and a precursor of an eventual transition to a solid state. An alternative way to probe that kind of physics is to use long-range interactions. Thus, for dipolar interactions the roton structure in the spectrum can be observed in the dilute and weakly-interacting limit \cite{Rotons}, signalling a possible transition to a supersolid \cite{FragmentedSS}.

The magnon dispersion \eqref{MagnonHighMomenta} does not develop a roton minimum upon increasing $na^3$. Rather, it flattens showing gradual increase of the quasiparticle mass. In terms of the above analogy with the Bose polaron, one can speak about \textit{magnon self-localization} [for the discussion of self-localization of polarons see Ref. \cite{Luis} and references therein]. In fact, cooperative self-localization of multiple impurities has been argued to represent the nucleation process for the phase separation transition \cite{Santamore}. On the other hand, one cannot exclude a possibility to find \textit{magneto-rotons} \cite{SpinRotons} in a more general case of unequal interactions, where the spin-wave dispersion becomes linear at the end-point. The resulting instability in this case could bring the system to a new phase, a \textit{magnon crystal}. Still retaining a uniform density, the mixed condensate would separate into an ordered array of domains characterized by alternating spin polarization \cite{MagnonCondensate}. Investigation of this intriguing possibility is the subject of on going work.  

\subsection{2D gas}

The $s$-wave scattering amplitude in 2D is given by \cite{Landau}
\begin{equation}
\label{f2D}
f_0(E_{\bm k})=\frac{2\pi}{\ln{(E_{\bm k}/E_a)}},
\end{equation}
where we have defined $E_a=\hbar^2/ma^2$. For a hard-core potential of the radius $R_e$ and at small momenta one has $f(\bm k^\prime,\bm k)\approx f_0(k)$ with $a=e^{\gamma} R_e/2$, where $\gamma\approx 0.577$ is the Euler-Mascheroni constant \cite{Schick}. The integral term on the r.h.s of Eq. \eqref{Vert3} cannot be ignored, and it defines the value of the chemical potential [the formula \eqref{mu}] via the transcendental equation 
\begin{equation}
\label{mu2D}
\mu=-\frac{2\hbar^2 n_0}{m}\frac{2\pi}{\ln p_c a},
\end{equation}
where $p_c\equiv\sqrt{2m\mu}/\hbar$ and one assumes $p_c a\ll 1$. Furthermore, by using the first formula in \eqref{SigmaViaTs} and assuming $\hbar\omega\approx\hbar^2 p^2/2m$, we obtain
\begin{equation}
\label{2Dcorrection}
\Sigma_s^{\uparrow\uparrow}(\mathsf p)-\Sigma_s^{\uparrow\downarrow}(\mathsf p)-\mu=-\frac{\pi\hbar^2 n_0}{2m}\frac{1}{\ln^2{p_c a}}\left(\frac{p}{p_c}\right)^2,
\end{equation}
which holds at $p\ll p_c$. Upon substitution into \eqref{Spec1} and comparison with the formula \eqref{magnonmass} [where we must let $n=n_0$] this yields
\begin{equation}
\label{2Ddrag}
\rho_{\uparrow\downarrow}=-\frac{1}{8\ln p_c a}mn,
\end{equation}
for the superfluid drag in 2D. The result agrees with that of Ref. \cite{Pastukhov2}. In contrast to the 3D case, here the drag contributes to the magnon dispersion already in the first order of the perturbation theory. This reflects the enhanced role of quantum fluctuations and polaronic effects in low dimensions.

Another important distinction from the 3D geometry is that the tail \eqref{2Dcorrection} cannot be reproduced by doing the Bogoliubov transformation of the Hamiltonian \eqref{DiluteHamiltonian}, in which $g_{\sigma\sigma^\prime}(\bm p, \bm q)$ is expressed via the 2D scattering length \eqref{f2D}. In this sense the concept of effective interaction does not apply here. Still, however, one can use the standard relationship \cite{Schick}
\begin{equation}
\label{g2D} 
g=-\frac{2\hbar^2}{m}f_0(2\mu)
\end{equation}
to calculate the chemical potential \eqref{mu2D} and the excitation spectrum without the entrainment.

For long-range dipolar interactions the formula \eqref{g2D} should be supplemented with the so-called anomalous term \cite{Landau}, which to the leading order depends linearly on the transferred momentum \cite{DipolarScattering},
\begin{equation}
g(\bm p, \bm q)=g-\frac{2\pi\hbar^2}{m} \lvert \bm p -\bm q \rvert r_\ast,
\end{equation}
where $r_\ast$ is the dipolar length. For $ng\ll \hbar^2/mr_\ast^2$ the phonon branch of the spectrum may develop a roton-maxon structure \cite{RotonSpectrum}. As regards the magnon dispersion, no traces of the dipolar tail remain since the transferred momentum is identically zero for the forward scattering, which defines the correction to the dispersion in this case [see Eq.\eqref{DiluteMagnon}]. In other words, contribution of the dipolar tail to the superfluid drag can be neglected in the first approximation \cite{footnote}.        

Finally, it is worth to point out, that the condition of weak interactions in 2D is automatically fulfilled in the range of validity of the formula \eqref{f2D}, i.e. $E_{\bm k}\ll E_a$. A different situation takes place in the vicinity of a shape resonance \cite{ResonantPairing}. Extension of our approach to this case will be given in a separate paper.

\section{Conclusions}

Our consideration generalizes the Beliaev diagrammatic theory to the case of a binary mixture of Bose-Einstein condensates. The elementary excitation spectrum consists of two gapless modes one of which takes the parabolic form \eqref{magnon} in the limit where the inter- and intra-species interactions are the same. We observe renormalization of the magnon mass due to the superfluid drag effect which contributes to the expansion of the kinetic energy of the system at small momenta. In the dilute regime the diagrams for the self-energy parts decouple into a set of independent ladders. This yields three effective potentials expressed via the corresponding scattering amplitudes. For weak interactions in 3D these potentials can be used to construct the effective Hamiltonian \eqref{DiluteHamiltonian} suitable for the perturbative expansion. The drag contributes to the magnon dispersion in the second order of the perturbation theory and can be calculated by dressing the magnons with the Bogoliubov phonon modes. The problem shares fruitful analogies with the physics of Bose polarons. Thus, an interesting direction for the future work is the search for \textit{magneto-rotons} and self-localized \textit{magnon crystals} in long-range interacting systems with specially designed microscopic potentials. In 2D we find renormalization of the magnon mass in the first approximation in $\beta$. This reflects the enhancement of quantum flcutuations in low dimensions. We thus expect the drag effect to play an important role in quantum-mechanical stabilization of a collapsing 2D Bose-Bose mixture already in the limit of weak interactions. 

\section{Acknowledgements}

S.V. acknowledges the support by the Government of the Russian Federation (Grant No. 074-U01) through ITMO Postdoctoral Fellowship scheme.                       

\section*{Appendix A: Bogoliubov transformation}

Consider the Hamiltonian \eqref{DiluteHamiltonian} and assume the effective interaction potential to be not dependent on the particle's spin and weak, i. e. $na^3\ll 1$. Following the standard procedure, we replace the operators $\hat{a}_{0, \sigma}$ and $\hat a^\dag_{0, \sigma}$ with c-numbers: $\hat{a}_{0, \sigma} = \sqrt{N_0}$. The occupation numbers for the states with finite momenta are assumed to be small. By retaining only quadratic terms in $\hat{a}_{\bm p, \sigma}$ and $\hat a^\dag_{\bm p, \sigma}$ with $\bm p \neq 0$ we get
\begin{widetext}
\begin{equation*}
 \begin{aligned}
\hat{H}_{\ast}=\sum_{{\bm p},\sigma}\varepsilon_{\bm p}^0 \hat{a}_{{\bm p},\sigma}^{\dag}    	
  \hat{a}_{{\bm p},\sigma}+
  \frac{n_0}{2} \sum_{{\bm p}} \left(  g(0,{\bm p}) (\hat{a}_{{\bm p},\sigma}\hat{a}_{\bm{-p},\sigma} +  \hat{a}_{{\bm p},\sigma}\hat{a}_{\bm{-p},\sigma^\prime} +\hat{a}_{\bm{-p},\sigma}\hat{a}_{{\bm p},\sigma^\prime}  + \hat{a}_{{\bm p},\sigma^\prime}\hat{a}_{\bm{-p},\sigma^\prime}) + h.c. \right) + \\ 
\frac{n_0}{2} \sum_{{\bm p}} g(\tfrac{{\bm p}}2,\tfrac{{\bm p}}2)( 4 \hat{a}_{{\bm p},\sigma}^{\dag} \hat{a}_{{\bm p},\sigma} + 4 \hat{a}_{{\bm p},\sigma^\prime}^{\dag} \hat{a}_{{\bm p},\sigma^\prime})
+\frac{n_0}{2} \sum_{{\bm p}} g(\tfrac{{\bm p}}2,\tfrac{\bm{-p}}2)( 2\hat{a}_{{\bm p},\sigma}^{\dag} \hat{a}_{{\bm p},\sigma^\prime} + 2\hat{a}_{{\bm p},\sigma^\prime}^{\dag} \hat{a}_{{\bm p},\sigma} + 2\hat{a}_{{\bm p},\sigma}^{\dag} \hat{a}_{{\bm p},\sigma} + 2 \hat{a}_{{\bm p},\sigma^\prime}^{\dag} \hat{a}_{{\bm p},\sigma^\prime} ) \\ - g(0,0) \sum_{{\bm p}} ( 4N \hat{a}_{{\bm p},\sigma}^{\dag} \hat{a}_{{\bm p},\sigma} + 4N \hat{a}_{{\bm p},\sigma^\prime}^{\dag} \hat{a}_{{\bm p}.\sigma^\prime} ),
 \end{aligned}
\end{equation*}
\end{widetext}

Denoting for simplicity $  \hat{a}_{{\bm p},\sigma} = \hat{a}_{{\bm p}},
  \; \hat{a}_{{\bm p},\sigma^\prime} = \hat{b}_{{\bm p}}
  \; $ and $ \; \varepsilon^\prime_{\bm p} = \varepsilon_{\bm p}^0 + 2 n_0 g(\tfrac{{\bm p}}2,\tfrac{{\bm p}}2) + n_0 g(\tfrac{{\bm p}}2,\tfrac{\bm{-p}}2) - 2 n_0 g(0,0)$ we rewrite the above equation in the form
  \begin{widetext}
\begin{equation*}
\hat{H}_{\ast}=\sum_{{\bm p}} \Big[ \varepsilon_{\bm p}^\prime \big( \hat{a}_{{\bm p}}^{\dag}
  \hat{a}_{{\bm p}} + \hat{b}_{{\bm p}}^{\dag}\hat{b}_{{\bm p}} \big) + 
  \frac{n_0}{2} \big( g(0,{\bm p}) (\hat{a}_{{\bm p}}\hat{a}_{\bm{-p}} +  \hat{a}_{{\bm p}}\hat{b}_{\bm{-p}} + \hat{a}_{\bm{-p}}\hat{b}_{{\bm p}}+ \hat{b}_{{\bm p}}\hat{b}_{\bm{-p}}) + h.c. \big)  + n_0 g(\tfrac{{\bm p}}2,\tfrac{\bm{-p}}2) \big( \hat{a}_{{\bm p}}^{\dag} \hat{b}_{{\bm p}} + \hat{b}_{{\bm p}}^{\dag} \hat{a}_{{\bm p}}\big) \Big].
\end{equation*}
\end{widetext}

Consider a unitary transformation $U$ with real coefficients and assume $\lambda_i(p), \beta_i(p)$ to be even functions of $p$.
\begin{eqnarray*}
 U \hat{a}_{{\bm p}} U^\dag = \lambda_1(p) \hat{a}_{{\bm p}} + \lambda_2(p) \hat{a}_{\bm{-p}}^\dag + \beta_1(p) \hat{b}_{{\bm p}} + \beta_2(p) \hat{b}_{{\bm p}}^\dag \\
U \hat{b}_{{\bm p}} U^\dag = \lambda_3(p) \hat{a}_{{\bm p}} + \lambda_4(p) \hat{a}_{\bm{-p}}^\dag + \beta_3(p) \hat{b}_{{\bm p}} + \beta_4(p) \hat{b}_{{\bm p}}^\dag 
\end{eqnarray*}

$U^\dag \hat{a}_{{\bm p}} U$ and $U^\dag \hat{b}_{{\bm p}} U$ can be understood as quasiparticle annihilation operators. We search for $U$ that diagonalises the Hamiltonian, so $U \hat{H}_{\ast} U^\dag  =  \sum_{{\bm p}}\omega_{\bm p}\big( \hat{a}_{{\bm p}}^{\dag} \hat{a}_{{\bm p}} + \hat{b}_{{\bm p}}^{\dag}\hat{b}_{{\bm p}} \big) + E_0$, where $\omega_{\bm p}$ is the excitation energy.

Using the relations $U[ \hat{H}_{\ast}, \hat{a}_{{\bm p}}^\dag] U^\dag =[ U \hat{H}_{\ast} U^\dag, U \hat{a}_{{\bm p}}^\dag U^\dag]$, $U[ \hat{H}_{\ast}, \hat{b}_{{\bm p}}^\dag] U^\dag =[ U \hat{H}_{\ast} U^\dag, U \hat{b}_{{\bm p}}^\dag U^\dag]$, we get the linear system 

\begin{widetext}
\begin{equation*}
\begin{split}
 \varepsilon_{\bm p}^\prime \big( \lambda_1 \hat{a}_{{\bm p}}^\dag + \lambda_2 \hat{a}_{\bm{-p}} +\beta_1\hat{b}_{{\bm p}}^\dag + \beta_2 \hat{b}_{\bm{-p}} \big)  + \frac{n_0}{2}  g(0,{\bm p}) \big(  (\lambda_1 + \lambda_3) \hat{a}_{\bm{-p}} + (\lambda_2 +\lambda_4) \hat{a}_{{\bm p}}^\dag + (\beta_1+\beta_3) \hat{b}_{\bm{-p}} + (\beta_2+\beta_4) \hat{b}_{{\bm p}}^\dag \big) + \\ + n_0 g(\tfrac{{\bm p}}2,\tfrac{\bm{-p}}2) \big( \lambda_3 \hat{a}_{{\bm p}}^\dag + \lambda_4 \hat{a}_{\bm{-p}} +\beta_3 \hat{b}_{{\bm p}}^\dag + \beta_4 \hat{b}_{\bm{-p}} \big) = \omega_{\bm p}\big( \lambda_1 \hat{a}_{{\bm p}}^\dag - \lambda_2 \hat{a}_{\bm{-p}} +\beta_1\hat{b}_{{\bm p}}^\dag - \beta_2 \hat{b}_{\bm{-p}} \big)
\end{split}
\end{equation*}

\begin{equation*}
\begin{split}
 \varepsilon_{\bm p}^\prime \big( \lambda_3 \hat{a}_{{\bm p}}^\dag + \lambda_4 \hat{a}_{\bm{-p}} +\beta_3 \hat{b}_{{\bm p}}^\dag + \beta_4 \hat{b}_{\bm{-p}} \big)  + \frac{n_0}{2}  g(0,{\bm p}) \big(  (\lambda_1 + \lambda_3) \hat{a}_{\bm{-p}} + (\lambda_2 +\lambda_4) \hat{a}_{{\bm p}}^\dag + (\beta_1+\beta_3) \hat{b}_{\bm{-p}} + (\beta_2+\beta_4) \hat{b}_{{\bm p}}^\dag \big) + \\ + n_0 g(\tfrac{{\bm p}}2,\tfrac{\bm{-p}}2) \big( \lambda_1 \hat{a}_{{\bm p}}^\dag + \lambda_2 \hat{a}_{\bm{-p}} +\beta_1\hat{b}_{{\bm p}}^\dag + \beta_2 \hat{b}_{\bm{-p}} \big) = \omega_{\bm p}\big( \lambda_3 \hat{a}_{{\bm p}}^\dag - \lambda_4 \hat{a}_{\bm{-p}} +\beta_3 \hat{b}_{{\bm p}}^\dag - \beta_4 \hat{b}_{\bm{-p}} \big)
\end{split}
\end{equation*}
\end{widetext}

The dispersion law can be obtained by equating the determinant to zero. We find
\begin{equation}
\label{gDiluteMagnon}
\omega_{\bm p} = \varepsilon_{\bm p}^0 + 2 n_0 \Big[ g(\tfrac{{\bm p}}2,\tfrac{{\bm p}}2) - g(0,0) \Big]
\end{equation}
in agreement with the formula \eqref{DiluteMagnon} and
\begin{widetext}
\begin{equation}
\label{gDilutePhonon}
\omega_{\bm p} =  \sqrt{(\varepsilon_{\bm p}^0)^2+4 n_0 \varepsilon_{\bm p}^0 
\big[ g(\tfrac{{\bm p}}2,\tfrac{\bm{-p}}2) + g(\tfrac{{\bm p}}2,\tfrac{{\bm p}}2) - g(0,0) \big] + 4 n_0^2 \Big[\big( g(\tfrac{{\bm p}}2,\tfrac{\bm{-p}}2) + g(\tfrac{{\bm p}}2,\tfrac{{\bm p}}2) - g(0,0) \big)^2 - g(0,{\bm p})^2 \Big]}, 
\end{equation}
\end{widetext}
which has the typical linear form at $\bm p\rightarrow 0$ and describes the excitation of phonons. Manifestation of the entrainment in this latter branch will be discussed in a separate paper.

\section*{Appendix B: General case of unequal masses, densities and interaction potentials}

In this section we consider a general situation - we have two types of bosons - ``a'' and ``b'' with different masses $m_a$ and $m_b$, correspondingly. Thus, we have two different bare Green's functions
\begin{eqnarray}
 \label{Ga1}
  G^{-1}_a(\omega, \textbf{p}) &=& \omega - \frac{p^2}{2m_a} +\mu_a +i0,  \\ \label{Ga2}
  G^{-1}_b(\omega, \textbf{p}) &=& \omega - \frac{p^2}{2m_b} +\mu_b +i0. 
\end{eqnarray}
Although now we have two different kind of particles the basic idea of the theory is the same. It is easy to show that the main contributions to the self-energy parts stem from the ladder diagrams shown in Fig. \ref{Vertex}, other contributions are small in the gas parameter. For simplicity we consider particle scattering amplitudes as momenta-independent, corresponding renormalized interaction vertexes are $g_{aa}, g_{ab}, g_{bb}$. We denote condensates densities as $n_a$ and $n_b$. So we have the following system of Dyson equations:
\begin{widetext}
\begin{equation}\label{Matr3}
  \left[
    \begin{array}{cccc}
      G^{-1}_a(\mathsf p) - \Sigma^{aa}_{11} & -(n_a n_b)^{1/2} g_{ab} & - n_a g_{aa} &  -(n_a n_b)^{1/2} g_{ab}\\
      -(n_a n_b)^{1/2} g_{ab} & G^{-1}_b(\mathsf p) - \Sigma^{bb}_{11} & -(n_a n_b)^{1/2} g_{ab} & - n_b g_{bb} \\
      n_a g_{aa} & -(n_a n_b)^{1/2} g_{ab} & G^{-1}_a(-\mathsf p) - \Sigma^{aa}_{11} & -\-(n_a n_b)^{1/2} g_{ab} \\
      -(n_a n_b)^{1/2} g_{ab} & - n_b g_{bb} & -(n_a n_b)^{1/2} g_{ab} & G^{-1}_b(-\mathsf p) - \Sigma^{bb}_{11} \\
    \end{array}
  \right] \left[
            \begin{array}{c}
              G_{aa}(\mathsf p) \\
              G_{ba}(\mathsf p) \\
              F^\dag_{aa}(\mathsf p) \\
              F^\dag_{ba}(\mathsf p) \\
            \end{array}
          \right]
          = \left[
              \begin{array}{c}
                1 \\
                0 \\
                0 \\
                0 \\
              \end{array}
            \right],
\end{equation}
\end{widetext}
where
\begin{eqnarray}
  \Sigma^{aa}_{11} &=& 2 n_a g_{aa} + n_b g_{ab}, \\
  \Sigma^{bb}_{11} &=& 2 n_b g_{bb} + n_a g_{ab},
\end{eqnarray}
and the same with a change of all indexes $a \leftrightarrow b$. First one should find the chemical potentials $\mu_a$ and $\mu_b$. We put $\mathsf p=0$ into \eqref{Matr3} and solve it. After cumbersome calculations we find a solution that provides poles in Green's functions at $\mathsf p=0$ and satisfies the condition $F^\dag_{aa} \approx -G_{aa}$ and $F^\dag_{bb} \approx -G_{bb}$:
\begin{eqnarray}
  \label{Chema}
  \mu_a &=& n_a g_{aa} + n_b g_{ab} ,\\ \label{Chemb}
  \mu_b &=& n_b g_{bb} + n_a g_{ab}.  
\end{eqnarray}
This equations generalize the corresponding equations for chemical potential in the main text. Now we can find Green's functions, all of them have the same denominator
\begin{widetext}
\begin{equation}\label{Den1}
  \begin{split}
    D(\omega, \textbf{p}) = & \omega^4 - \left[ \varepsilon^2_a(p) + 2 n_a g_{aa} \varepsilon_a(p) + \varepsilon^2_b(p) + 2 n_b g_{bb} \varepsilon_b(p) \right] \omega^2 + \\
    + &  \varepsilon_a(p) \varepsilon_b(p) \left[ \varepsilon_a(p) \varepsilon_b(p) + 2 n_b g_{bb} \varepsilon_a(p) + 2 n_a g_{aa} \varepsilon_b(p) - 4 n_a n_b (g^2_{ab} - g_{aa} g_{bb})  \right].
  \end{split}
\end{equation}
\end{widetext}
Using this formula we can find quasiparticles spectra in the system. One has a positive root if $g^2_{ab}<g_{aa} g_{bb}$, which corresponds to the miscibility condition \eqref{miscibility}. After some calculations we get
\begin{widetext}
\begin{equation}\label{Spec2}
  \begin{split}
    \omega^2(\textbf{p})= & \frac{1}{2} \Biggl[ \varepsilon^2_a(p) + 2 n_a g_{aa} \varepsilon_a(p) + \varepsilon^2_b(p) + 2 n_b g_{bb} \varepsilon_b(p) \pm \\ \pm & \sqrt{\left(\varepsilon^2_a(p) + 2 n_a g_{aa} \varepsilon_a(p) - \varepsilon^2_b(p) - 2 n_b g_{bb} \varepsilon_b(p)\right)^2 + 16 n_a n_b g^2_{ab} \varepsilon_a(p) \varepsilon_b(p)  } \Biggr].  \\
  \end{split}
\end{equation}
\end{widetext}
One can see that if $g_{ab}=0$ then we get a usual phonon spectra for ``a'' and ``b'' particles, if we consider symmetric case then we get spectra given in the main text. 

The Green's functions have the following form
\begin{widetext}
\begin{eqnarray}
  G_{aa}(\omega, \textbf{p}) &=& \Bigl[ \omega^3 + \left(\varepsilon_a(p) + n_a g_{aa} \right) \omega^2 -  \varepsilon_b(p) \left(\varepsilon_b(p) + 2 n_b g_{bb}\right) \omega - \\ \nonumber &-& \varepsilon_b(p) \left(\varepsilon_b(p)\left[\varepsilon_a(p) + n_a g_{aa}\right] + 2 \left[g_{bb}\varepsilon_a(p) - n_a g^2_{ab} + n_a g_{aa} g_{bb}\right]n_b \right) \Bigr]/D(\omega, \textbf{p}), \\
  G_{ba}(\omega, \textbf{p}) &=& \frac{g_{ab}\sqrt{n_a n_b}(\omega+\varepsilon_a(p))(\omega+\varepsilon_b(p))}{D(\omega, \textbf{p})}, \\
  F\dag_{aa}(\omega, \textbf{p}) &=& \frac{n_a \varepsilon_b(p) \left[ g_{aa}\varepsilon_b(p) - 2 n_b g^2_{ab} + 2 n_b g_{aa} g_{bb} \right] - n_a g_{aa} \omega^2 }{D(\omega, \textbf{p})}, \\
  F^\dag_{ba}(\omega, \textbf{p}) &=& \frac{g_{ab}\sqrt{n_a n_b}(\omega+\varepsilon_a(p))(\varepsilon_b(p)-\omega)}{D(\omega, \textbf{p})},
\end{eqnarray}
\end{widetext}
other Green's functions can be obtained by changing $a \leftrightarrow b$.

\section*{Appendix C: Experimental detection of magnons}

The spin-wave dispersion can be extracted from the measurements of the dynamic structure factor
\begin{equation}
S_\mathrm{m}(\bm q,\omega)=\frac{1}{n}\int <\hat n_\uparrow (\bm r,t)\hat n_\downarrow(0,0)>e^{-(i\bm q\bm r-\omega t)}d\bm r dt
\end{equation}
as detailed in Ref. \cite{Carusotto}. Difficulties may arise in the case of the parabolic dependence \eqref{magnon} since the spectrum takes this form at the miscibility transition point, where the condensates tend to separate. What could be easier observed in this case is a change in the  static structure factor
\begin{equation}
S_\mathrm{m}(\bm q)=\frac{<\hat n_{\uparrow,\bm q}\hat n_{\downarrow,-\bm q}>}{N},
\end{equation}
where $\hat n_{\sigma,\bm q}=\int \hat n_\sigma(\bm r) e^{-i\bm q\bm r}d\bm r$. Within the Bogoliubov approach one has $\hat n_{\sigma,\bm q}=\sqrt{N}(\hat a_{\sigma,-\bm q}^\dagger+\hat a_{\sigma,\bm q})$. By substituting the Bogoliubov transformation for the operators $a_{\sigma,\bm q}$ and taking advantage of the fact that at $T=0$ one has $<\hat b_{\bm q}\hat b_{\bm q}^\dagger>=1$ we find at $q\rightarrow 0$
\begin{equation}
S_\mathrm{m}(\bm q)=\frac{1}{4}\sqrt{\frac{\hbar^2 p^2}{mn}}\frac{\sqrt{g_a}-\sqrt{g_s}}{\sqrt{g_a g_s}},
\end{equation}
where $g_{s,a}=g_{\uparrow\uparrow}\pm g_{\uparrow\downarrow}$. One can see that the static structure factor diverges like $\sim 1/\sqrt{g_a}$ as $g_{\uparrow\uparrow}\rightarrow g_{\uparrow\downarrow}$.

To study renormalization of the magnon mass due to the entrainment one can employ the experimental scheme discussed in Ref. \cite{Marti}. In this experiment a standing wave of magnons is imprinted onto the condensate by illuminating the atoms with two equal-frequency circularly polarized light beams and modulating their intensity at the frequency corresponding to a Raman transfer between Zeeman levels. The dispersion relation then can be obtained by analyzing the dynamics of the resulting spin distribution. Interestingly, the authors ascertain a tiny increase of the magnon mass as compared to the bare mass of atoms.


\begin{thebibliography}{99}

\bibitem{Fetter} A. L. Fetter and J. D. Walecka, \textit{Quantum Theory of Many-Particle System}, Dover Publications, New York (2003).

\bibitem{Friedel} M. F. Crommie, C. P. Lutz and D. M. Eigler, Nature \textbf{363}, 524 (1993).

\bibitem{Kirzhnits} D. A. Kirzhnits and Yu. A. Nepomnyashchii, Sov. Phys.
JETP \textbf{32}, 1191 (1971).

\bibitem{Nepomnyashchii} Yu. A. Nepomnyashchii, Theor. Math. Phys. \textbf{8}, 928 (1971).

\bibitem{Rosenzweig} H. Kadau, M. Schmitt, M. Wenzel, C. Wink, T. Maier, I. Ferrier-Barbut, and T. Pfau, Nature (London) \textbf{530}, 194 (2016).

\bibitem{Rotons} L. Chomaz, R. M. W. van Bijnen, D. Petter, G. Faraoni, S. Baier, J. H. Becher, M. J. Mark, F. Wachtler, L. Santos, F. Ferlaino, Nature Physics (2018).

\bibitem{FragmentedSS} S. V. Andreev, Phys. Rev. B \textbf{95}, 184519 (2017).

\bibitem{Wigner} E. Wigner, Phys. Rev. \textbf{46}, 1002 (1934).

\bibitem{RotonSpectrum} A. Boudjemaa and G. V. Shlyapnikov, Phys. Rev. A \textbf{87}, 025601 (2013).

\bibitem{1Dscattering} S. V. Andreev, Phys. Rev. B \textbf{92}, 041117(R) (2015).

\bibitem{DipolarScattering} M. A. Baranov, A. Micheli, S. Ronen, and P. Zoller, Phys. Rev. A \textbf{83}, 043602 (2011); J. Levinsen, N. R. Cooper, and G. V. Shlyapnikov, Phys. Rev. A \textbf{84}, 013603 (2011).

\bibitem{Beliaev} S. T. Beliaev, Zh. Eksp. Teor. Fiz. \textbf{34}, 417 (1958) [Sov. Phys. JETP \textbf{7}, 289 (1958)].

\bibitem{PancakeRotons} L. Santos, G. V. Shlyapnikov, and M. Lewenstein, Phys. Rev. Lett. \textbf{90}, 250403 (2003).

\bibitem{Review} M. A. Baranov, Phys. Rep. \textbf{464}, 71 (2008).

\bibitem{Filaments} I. Ferrier-Barbut, H. Kadau, M. Schmitt, M. Wenzel, and T. Pfau, Phys. Rev. Lett. \textbf{116}, 215301 (2016).

\bibitem{Pfau} T. Pfau at BEC 2017, Spain (2017).

\bibitem{Fluctuations} F. Wachtler and L. Santos, Phys. Rev. A \textbf{93}, 061603(R) (2016).

\bibitem{ErbiumDrop} L. Chomaz, S. Baier, D. Petter, M. J. Mark, F. Wachtler, L. Santos, and F. Ferlaino
Phys. Rev. X \textbf{6}, 041039 (2016).

\bibitem{DysprosiumDrop} M. Schmitt, M. Wenzel, F. Bottcher, I. Ferrier-Barbut, T. Pfau, Nature (London) \textbf{539}, 259 (2016).

\bibitem{DipolarDropletsScience} Igor Ferrier-Barbut and Tilman Pfau, Science \textbf{359}, 274 (2018).

\bibitem{DiluteSupersolid} Z.-K. Lu, Y. Li, D. S. Petrov, and G. V. Shlyapnikov, Phys. Rev. Lett. \textbf{115}, 075303 (2015).

\bibitem{PolarMolecules} On recent advances in creation and manipulation of polar atomic molecules see T. Takekoshi, L. Reichsollner, A. Schindewolf, J. M. Hutson, C. R. L. Sueur, O. Dulieu, F. Ferlaino, R. Grimm, and H.-C. Nagerl, Phys. Rev. Lett. \textbf{113}, 205301 (2014); Sebastian A. Will, Jee Woo Park, Zoe Z. Yan, Huanqian Loh, and Martin W. Zwierlein, Phys. Rev. Lett. \textbf{116}, 225306 (2016). 

\bibitem{Petrov2014} D. S. Petrov, Phys. Rev. Lett. \textbf{112}, 103201 (2014).

\bibitem{Will} Private communication from Sebastian Will.

\bibitem{MOES} L. V. Butov, A. C. Gossard, and D. S. Chemla, Nature (London) \textbf{418}, 751 (2002).

\bibitem{ResonantPairing} S. V. Andreev, Phys. Rev. B \textbf{94}, 140501(R) (2016).

\bibitem{Petrov2015} D. S. Petrov, Phys. Rev. Lett. \textbf{115}, 155302 (2015).

\bibitem{BoseMixture} C. R. Cabrera, L. Tanzi, J. Sanz, B. Naylor, P. Thomas, P. Cheiney and L. Tarruell, Science \textbf{359}, 301 (2018); G. Semeghini, G. Ferioli, L. Masi, C. Mazzinghi, L. Wolswijk, F. Minardi, M. Modugno, G. Modugno, M. Inguscio and M. Fattori, arXiv:1710.10890 (2017).

\bibitem{Petrov2016} D. S. Petrov and G. E. Astrakharchik, Phys. Rev. Lett. \textbf{117}, 100401 (2016).

\bibitem{PetrovNature} D. S. Petrov, \textit{Liquid beyond the van der Waals paradigm}, Nature (2018).

\bibitem{AB} A. F. Andreev and E. P. Bashkin, Sov. Phys. JETP \textbf{42}, 164 (1976).

\bibitem{Goldstein} Elena V. Goldstein and Pierre Meystre, Phys. Rev. A \textbf{55}, 2935 (1997).

\bibitem{Berman} C. P. Search, A. G. Rojo, and P. R. Berman, Phys. Rev. A \textbf{64}, 013615 (2001).

\bibitem{Alexandrov} A. S. Alexandrov and V. V. Kabanov, J. Phys.: Condens. Matter \textbf{14}, L327 (2002).

\bibitem{Eckardt} A. Eckardt, C. Weiss, and M. Holthaus, Phys. Rev. A \textbf{70}, 043615 (2004).

\bibitem{Christensen} R. S. Christensen, J. Levinsen and G. M. Bruun, Phys. Rev. Lett. \textbf{115}, 160401 (2015). 

\bibitem{Pitaevskii} E. M. Lifshitz and L. P. Pitaevskii, \textit{Statistical Physics, Part 2} (Pergamon Press, Oxford, 1980).

\bibitem{AGD} A. A. Abrikosov, L. P. Gorkov, and I. E. Dzialoshinski, \textit{Methods of Quantum Field Theory in Statistical Physics} (Dover, New York, 1975).

\bibitem{Gavoret} G. Gavoret and P. Nozieres, Ann. Phys. \textbf{28}, 349 (1964).

\bibitem{ABnote} The quantity $\rho_{\uparrow\downarrow}$ enters the expansion of the ground state energy in terms of the superfluid velocities of the components $\delta E=\frac{1}{2}\sum\limits_{\sigma\sigma^\prime} \int \rho_{\sigma\sigma^\prime} \bm v_\sigma \bm v_{\sigma^\prime} d\bm x$.
    
\bibitem{Halperin1} B. I. Halperin, Phys. Rev. B \textbf{11}, 178 (1975).

\bibitem{Halperin2} B. I. Halperin and P. C. Hohenberg, Phys. Rev. \textbf{188}, 898 (1969).

\bibitem{Nespolo} Jacopo Nespolo \textit{et. al.}, New J. Phys. \textbf{19}, 125005 (2017).

\bibitem{Lozovik} Yu. E. Lozovik and V. I. Yudson, Physica (Amsterdam) \textbf{93}A, 493 (1978).

\bibitem{Taylor} J. R. Taylor, \textit{Scattering Theory} (John Wiley \& Sons, New York, 1972).

\bibitem{Pastukhov1} V. Pastukhov, Phys. Rev. A \textbf{95}, 023614 (2017).

\bibitem{Fil} D. V. Fil and S. I. Shevchenko, Phys. Rev. A \textbf{72} 013616 (2005).

\bibitem{Luis} L. A. Pena Ardila and S. Giorgini
Phys. Rev. A \textbf{92}, 033612 (2015).

\bibitem{Santamore} D. H. Santamore and E. Timmermans, New Journal of Physics \textbf{13}, 103029 (2011).  

\bibitem{SpinRotons} H. A. Mook, N. Wakabayashi, and D. Pan, Phys. Rev. Lett. \textbf{34}, 1029 (1975); H. A. Mook and C. C. Tsuei, Phys. Rev. B \textbf{16}, 2184 (1977); T. Kaneyoshi, J. Phys. Soc. Jpn. \textbf{45} 1835 (1978);  V.A. Singh, L.M. Roth, J. Appl. Phys. \textbf{49}, 1642 (1978).

\bibitem{MagnonCondensate} It is worth to mention the experimental work [F. Fang, R. Olf, S. Wu, H. Kadau and D. M. Stamper-Kurn, Phys.Rev. Lett. \textbf{116}, 095301 (2016)], where macroscopic polarization textures were observed upon cooling a magnon gas. 

\bibitem{Landau} L.D. Landau and E.M. Lifshitz, \textit{Quantum Mechanics}  (Butterworth-Heinemann, Oxford, 1999).

\bibitem{Schick} M. Schick, Phys. Rev. A \textbf{3}, 1067 (1971).

\bibitem{Pastukhov2} P. Konietin and V. Pastukhov, J. Low Temp. Phys. \textbf{190}, 256 (2018).

\bibitem{footnote} In fact, it enters the expression \eqref{2Ddrag} implicitly via the 2D scattering length $a$. Thus, one has $a\sim r_\ast$ for a purely dipolar potential.

\bibitem{Carusotto} I. Carusotto, J. Phys. B: At. Mol. Opt. Phys. \textbf{39}, S211 (2006).

\bibitem{Marti} G. E. Marti, A. MacRae, R. Olf, S. Lourette, F. Fang, and D. M. Stamper-Kurn, Phys. Rev. Lett. \textbf{113}, 155302 (2014).

\end{thebibliography}
\end{document}